\documentclass[prb,twocolumn,amsmath,amssymb,citeautoscript]{revtex4}
\usepackage{times}
\usepackage{graphicx}
\usepackage{dcolumn}
\usepackage{bm}

\begin{document}

\title{
Correlation Induced Inhomogeneity in Circular Quantum Dots
}

\author{Amit Ghosal,$^{1}$ A.~D.~G\"u\c{c}l\"u,$^2$ C.~J.~Umrigar,$^2$ Denis Ullmo,$^{1,3}$ and Harold U. Baranger$^1$}


\affiliation{$^1$Department of Physics, Duke University,
Durham, North Carolina 27708-0305}

\affiliation{$^2$Theory Center and Laboratory of Atomic and Solid State Physics, Cornell University,
Ithaca, New York 14853}

\affiliation{$^3$CNRS; Univ. Paris-Sud, LPTMS UMR 8626, 91405 Orsay Cedex France}

\date{\bf Published 23 April 2006; Nature Physics 2, 336-340 (2006)}


\maketitle

{\bf Properties of the ``electron gas'' -- in which conduction electrons interact by means of Coulomb forces but ionic potentials are neglected -- change dramatically depending on the balance between kinetic energy and Coulomb repulsion. The limits are well understood~\cite{ElectronLiqBook}: For very weak interactions (high density), the system behaves as a Fermi liquid, with delocalized electrons. In contrast, in the strongly interacting limit (low density), the electrons localize and order in a Wigner crystal. The physics at intermediate densities remains a subject of fundamental research\cite{Tanatar,Italian2DEG,metal-ins,sudip,waintal}. Here, we study the intermediate-density electron gas confined to a circular disk, where the degree of confinement can be tuned to control the density. Using accurate quantum Monte Carlo techniques~\cite{Foulkes}, we show that the electron-electron correlation induced by an increase of the interaction first smoothly causes rings, and then angular modulation, without any signature of a sharp transition in this density range. We conclude that inhomogeneities in a confined system, which exist even without interactions, are significantly enhanced by correlations.}

Quantum dots~\cite{Tarucha} -- a nanoscale island containing a puddle of electrons -- provide a highly tunable and simple setting to study the effects of large Coulomb interaction. They introduce level quantization and quantum interference in a controlled way, and can in principle be made in the very low density regime where correlation effects are strong \cite{Reuter05}. There are in addition natural parallels between quantum dots and other confined systems of interacting particles, such as cold atoms in traps.

We therefore consider a model quantum dot consisting of electrons moving in a two-dimensional (2D) plane, with kinetic energy $(-\frac{1}{2}\sum_i \nabla_i^2)$, and interacting with each other by long-range Coulomb repulsion ($\sum_{i<j}|{\bf r}_i - {\bf r}_j|^{-1}$). All energies are expressed in atomic units, defined by $\hbar \!=\! e^2/\epsilon \!=\! m^* \!=\! 1$ (with electronic charge $e$, effective mass $m^*$, and dielectric constant $\epsilon$). The electrons are confined by an external quadratic potential $V_{\rm ext}({\bf r})=\frac{1}{2}\omega^2 r^2$ with circular symmetry and spring constant $\omega$. The ratio between the strength of the Coulomb interaction and the kinetic energy is usually characterized by the interaction parameter $r_s \equiv (\pi n)^{-1/2}$, with $n$ the density of electrons. For our confined system in which $n({\bf r})$ varies, we define $r_s$ in the same way using the mean density $\bar{n} \equiv \int n^2({\bf r})d{\bf r}/N$. We have studied this system up to $N=20$ electrons. The spring constant $\omega$ makes the oscillator potential narrow (for large $\omega$) or shallow (for small $\omega$); it thereby tunes the average density of electrons between high and low values, thus controlling $r_s$. For example, for $N=20$, varying $\omega$ between $3$ and $0.0075$ changes $r_s$ from $0.4$ to $17.7$. The radius of the dot grows significantly as $r_s$ increases, in an approximately linear fashion (see Fig. 1).

In the bulk 2D electron gas, numerical work suggests a transition from a Fermi liquid state to a Wigner crystal around $r_s^c \approx$ 30-35 ~\cite{Tanatar,Italian2DEG,waintal}. On the other hand, experiments on the 2D electron gas (which include, of course, disorder and residual effects of the ions) observe more complex behavior, including evidence for a metal-insulator transition~\cite{metal-ins}.

Circular quantum dots have been studied previously using a variety of methods, yielding a largely inconclusive scenario. Many studies \cite{ReimannRMP,Grabert,YannLand} have used density functional theory or Hartree-Fock. These typically predict charge or spin density wave order even for modest $r_s$ (unless the symmetry is restored after the fact \cite{YannLand}) which are thought to be unphysical. Exact diagonalization calculations~\cite{ReimannED,CIpaper} can be highly accurate but are restricted to small $N$ and $r_s$. Path integral quantum Monte Carlo (PIMC) has also been applied: One study~\cite{Egger} found a crossover from Fermi liquid to ``Wigner molecule'' at $r_s \approx 4$ -- a value significantly smaller than the 2D bulk $r_s^c$. Another study \cite{Filinov}, using different criteria, found a two-stage transition for $r_s$ larger than $r_s^c$. Although PIMC treats interactions accurately, it has its own systematic and statistical problems; for instance, it generates a thermal average of states with different $L$ and $S$ quantum numbers, preserving only $S_z$ symmetry.
To avoid these various difficulties and so clarify the scenario, we have carried out a study using the variational Monte Carlo (VMC) and diffusion Monte Carlo (DMC) techniques~\cite{Foulkes}, which we used previously to study both circular \cite{Pederiva,Devrim} and irregular \cite{Ghosal} dots at $r_s \sim 2$. This method is free of the problems of PIMC but is approximate in that a ``fixed-node'' error is made. We believe the latter is small for the range of parameters studied here (see the ``Methods'' section and the detailed comparison in the supplementary material).

\begin{figure}[t]
\includegraphics[width=3.375in,clip]{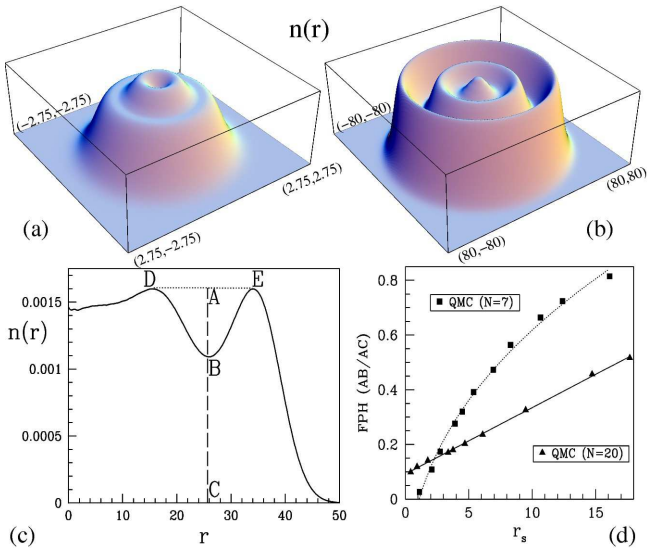}
\caption{
Electron density, $n({\bf r})$, for the ground state of a $N=20$ circular quantum dot ($L=0$, $S=0$). The extrapolated quantum Monte Carlo (QMC) estimator is used \cite{Foulkes}. 
\textbf{(a)}~High density: $r_s \approx 0.4$ ($\omega=3.0$). 
\textbf{(b)}~Low density: $r_s \approx 15$ ($\omega=0.01$). Note the dramatic change in density profile upon increasing $r_s$: the electron-electron correlation caused by stronger interactions at low density produces sharp rings. The 3 ring structure agrees with that seen in the classical limit. 
Notice the significant increase in the radius of the dot for larger $r_s$:
it changes from approximately $2.75$ to $80$ (in atomic units)
by increasing $r_s$ from $0.4$ to $15$.
\textbf{(c)}~Radial cut of $n({\bf r})$ for $r_s \approx 10$ ($\omega=0.02$) where the 3-ring structure is about to appear. The modulation is quantitatively characterized by the fractional peak height (FPH): draw the line tangential to the two outer peaks of $n({\bf r})$ ($\overline{DE}$), then find the vertical line $\overline{AC}$ along which the distance from $\overline{DE}$ to $n({\bf r})$ is maximum, and finally define the FPH as the ratio of the two lengths $\overline{AB}/\overline{AC}$.
\textbf{(d)}~FPH as a function of $r_s$ for $N=20$ and $7$. The curve for
$N=20$ is linear and completely featureless for $r_s \lesssim 18$. The solid line is a linear fit to the data. For smaller $N$, radial modulation in $n({\bf r})$ becomes stronger, leading to $FPH \rightarrow 1$ for large $r_s$, and a deviation from linearity occurs (the dotted line $\sim r_s^{0.41}$ is the best fit for $N=7$). For our largest $r_s$, the FPH typically grows with decreasing $N$, though not always monotonically. For example, it is largest for $N=7$, which yields a `perfect crystal' with equidistant electrons and thus produces a peak in the addition energy [see Fig. 3(a)]. The Monte Carlo statistical error is less than the size of the points.
}
\label{fig:Fig1}
\end{figure}

Results for the electron density, $n({\bf r})$, are shown in Fig.\,\ref{fig:Fig1}. There is a dramatic change in $n({\bf r})$ upon increasing interaction strength: For weak interactions [panel (a)], the density is rather homogeneous; the small modulation seen is caused by shell effects in the orbitals of the mean-field problem. In contrast, large $r_s$ induces strong radial modulation in $n({\bf r})$ [panel (b)], resulting in the formation of rings. 
Interestingly, for $r_s > 10$ the number of rings for each $N$ is the same as that seen in the classical limit~\cite{Bedanov,Shklovskii} ($r_s \rightarrow \infty$), e.g. three rings for $N=20$.
In all the cases we consider, the density $n({\bf r})$ is circularly
symmetric, as is the density of spin up and down electrons separately, as
required in two-dimensional systems, since we work with states of definite angular momentum $L$.

We find that the formation of rings and the increase in their sharpness is completely \textit{smooth}. This is shown quantitatively in Fig.\,\ref{fig:Fig1}(d) by using the fractional peak height [FPH, defined in Fig.\,\ref{fig:Fig1}(c)] of the outer ring to characterize the degree of structure. In the resulting curve for FPH as a function of $r_s$, no deviations or special value of $r_s$ can be seen.

Having established the role of strong correlations in the formation of radial rings,
which may be detectable in scanning probe measurements, we turn to angular modulation -- the issue of correlation induced localization of the individual electrons in each of the circular rings. We therefore consider the pair-densities $g_{\sigma\sigma'}({\bf r}_0;{\bf r})$ -- the probability of finding an electron with spin $\sigma'$ at location ${\bf r}$ when an electron with spin $\sigma$ is held fixed at ${\bf r}_0$ -- and $g_T = g_{\uparrow\uparrow} + g_{\uparrow\downarrow}$. These detect, in addition to radial rings, any angular structure induced by the interactions.

The most prominent feature in the pair density is a hole around the location of the fixed electron. For unlike spins, it is caused purely by Coulomb repulsion (correlation hole), while for like spins the antisymmetry of the wave function plays an important role (exchange hole). For small $r_s$, correlation is weak, so the hole in $g_{\sigma,-\sigma}$ is much smaller than that in $g_{\sigma,\sigma}$. As $r_s$ increases, the correlation hole grows bigger, becoming of the same size as the exchange hole around $r_s \approx 4$-$5$.

\begin{figure}[t]
\includegraphics[width=3.375in,clip]{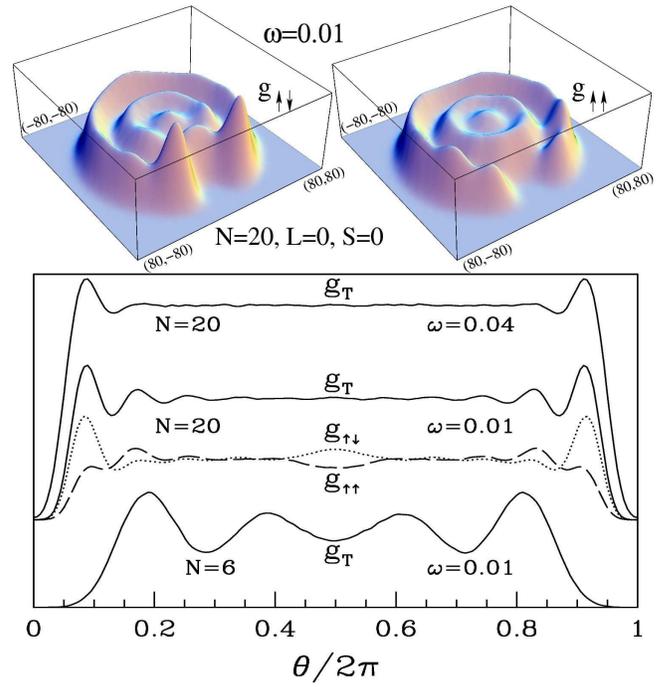}
\caption{
Pair density of the circular quantum dot with an up-electron fixed on the outer ring.
\textbf{Top:} $g_{\uparrow\downarrow}({\bf r}_0;{\bf r})$ (left) and $g_{\uparrow\uparrow}({\bf r}_0;{\bf r})$ (right) for the $N=20$ ground state ($L=0$, $S=0$) with $r_s \approx 15$ ($\omega=0.01$, ${\bf r}_0=(57,0)$). Short-range order develops near the fixed electron, indicating ``incipient'' Wigner localization but not true long-range order. 
\textbf{Bottom:} Evolution of angular oscillations along the outer ring with $r_s$ and $N$. 
The top trace shows $0.65 \times g_T$ for $r_s \approx 6$, $N=20$: though strong radial modulation has already appeared, leading to ``ring formation'', there is almost no angular modulation.
The middle trace is $g_T$ for $r_s \approx 15$, $N=20$: clear angular structure is present, though compared to the ring modulation it is weak and short-range. Spin-resolved angular structure is also shown here; note the peculiar bump at $\theta = \pi$.
The bottom trace is $g_T$ for $r_s \approx 16$, $N=6$ ($L=0$, $S=0$): for small $N$, angular modulation is clearly stronger.
(The y-axis is shifted and scaled for $N=20$ for clarity.)
}
\label{fig:Fig2}
\end{figure}

Results for the pair density in the circular quantum dot are shown in Fig.\,\ref{fig:Fig2} for an up-electron fixed on the outer ring. For $N=20$ at large $r_s$, there are clear oscillations along the angular direction near ${\bf r}_0$. This signals ``incipient'' Wigner localization. However, these oscillations are weak (weaker than the radial modulation) and short-ranged (damped), indicating that long-range order is not yet established.  As for radial modulation, the amplitude of the angular oscillations grows continuously, without any threshold value.

The evolution of the angular oscillations as a function of $r_s$ and $N$ is illustrated in the bottom panel of Fig. 2. Comparing the top two traces, for $r_s \approx 6$ and $15$ at $N=20$, we see that $g_{T}$ is almost featureless even for an $r_s$ substantially bigger than 1, while short-range oscillations have set in by our largest $r_s$. The weakness of these oscillations suggests that electrons remain more or less delocalized along the ring for $N=20$ up to the largest $r_s$ studied. An intriguing feature of the spin-resolved pair densities shown is the bump at $\theta=\pi$: $g_{\uparrow\uparrow}$ decreases while $g_{\uparrow\downarrow}$ increases. This feature is present for all $r_s \geq 4$ and grows with increasing interaction strength; we have no explanation for it at this time. Turning now to smaller $N$, we find that two rings are present for $N=6$ at large $r_s$: the outer one has $5$ electrons while the remaining electron is at the center. The lower trace in Fig. \ref{fig:Fig2} shows that individual electrons are better localized for small $N$, a behavior that we find holds quite generally.

\begin{figure}[t]
\includegraphics[width=3.375in,clip]{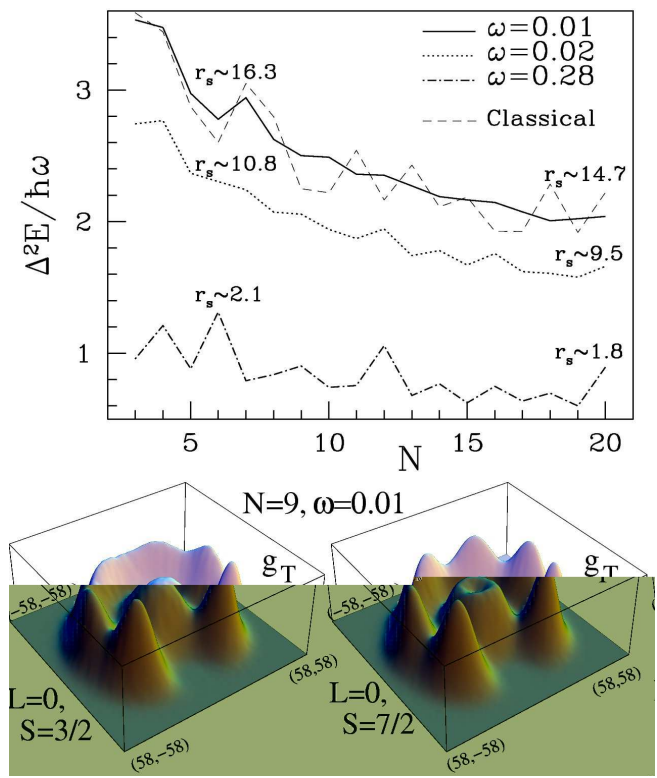}
\caption{
Ground state energy.
\textbf{Top:} Addition energy (normalized) as a function of $N$ for three different $\omega$ and for the classical limit \cite{Bedanov} $r_s \rightarrow \infty$. As interactions strengthen because of decreasing $\omega$, the mesoscopic fluctuations in $\Delta^2E$ become weaker. Note that this happens more readily in the small $N$ limit. Features in the $\omega=0.01$ trace at small $N$ are remarkably similar to those found in the classical limit, showing that electrons are nearly localized for small $N$. (The zero of the y-axis is offset for clarity, and the normalization of the classical trace is arbitrary.)
\textbf{Bottom:} $g_T$ for $N=9$, $r_s \approx 15$ ($\omega=0.01$), keeping an electron fixed on the outer ring (${\bf r}_0 \approx (37,0)$), for $L=0$, $S=3/2$ (left, the usual ground state) and $L=0$, $S=7/2$ (right). Increase of $r_s$, particularly for small $N$, often brings a strongly polarized state very close in energy to the ``usual'' ground state. In the case shown, the two states become essentially degenerate ($E=1.464651 H^*$) within statistical errors. The extent of Wigner localization is clearly stronger for the $S=7/2$ state.
}
\label{fig:Fig3}
\end{figure}

We next turn our attention to the addition energy, $\Delta^2E(N) = E_G(N \!+\!1) + E_G(N\!-\!1) - 2 E_G(N)$, where $E_G(N)$ is the ground state energy of the dot with $N$ electrons. This is experimentally accessible as the spacing between conductance peaks in a Coulomb blockade transport measurement, and is given by the charging energy in the simplest model of a quantum dot \cite{Tarucha}. Our results for $\Delta^2E(N)$ (normalized by $\omega$) for different interaction strengths are shown in Fig.\,3 ($r_s$ for fixed $\omega$ varies slightly with $N$). For $r_s \approx 2$, $\Delta^2E (N)$ is similar to previous studies~\cite{ReimannRMP,Pederiva}: non-interacting ``shell effects'' produce strong peaks for closed shell configurations ($N=6$, $12$, $20$). At larger $r_s$, the peaks weaken considerably, reducing mesoscopic fluctuations in $\Delta^2E$. For similar $r_s$, shell effects are more strongly affected for small $N$, while their remnant persists for large $N$. For comparison, we plot the addition energy in the classical limit \cite{Bedanov,Shklovskii} obtained from the ground state energies in Ref.\,\onlinecite{Bedanov}. The remarkable similarity to our quantum result for small $N$ at the largest $r_s$ is strong evidence for electron localization.

Strong correlations can shuffle the energy ordering of different quantum
states at fixed $N$. However, for $\omega \!>\! 0.01$, the ground state remains consistent with  Hund's first rule (except for $N=3$). For smaller $\omega$, we see a tendency toward violation of this rule, primarily for small $N$ (which are in general more affected by strong correlations), as in the following example. For $N = 9$, the Hund's rule ground state has $(L,S) \!=\! (0,3/2)$. We find that for $\omega \!=\! 0.01$, the highly polarized state $(0,7/2)$ becomes degenerate with the usual ground state (within our numerical accuracy). (All other $(L,S)$ states lie higher in energy.) Note that $S \!=\! 7/2$ requires promotion between non-interacting shells, and so lies much higher in energy in the weakly interacting limit. At large $r_s$, this difference is overcome by the gain in interaction energy. The pair density $g_T$ for both these $N \!=\! 9$ states is shown in Fig.~3. The more polarized state is clearly more localized; we find this is generally the case, as expected, since exchange keeps the electrons apart. 

\textit{In conclusion, the scenario that emerges here is significantly different from that for the bulk. The gradual emergence of the radial oscillations is connected to the fact that the translational symmetry is necessarily broken, and so the interactions can readily amplify existing inhomogeneities. The development of the addition energy curves further supports this point: the structure caused by quantum interference is rapidly suppressed by the interactions leading to surprisingly smooth behavior over a wide range. Thus, strong correlations and incipient localization should be taken into account for a very broad range of interaction strength.}

\bigskip
\noindent\textbf{Methods:}

As a starting point, we use the Kohn-Sham orbitals obtained from a density functional calculation done in the local density approximation. We then perform a variational Monte Carlo (VMC) calculation using a trial wave function, $\Psi_T$, which is a linear combination of products of up- and down- spin Slater determinants  of the Kohn-Sham orbitals multiplied by a Jastrow factor. The Jastrow factor effectively describes the dynamic correlation between the electrons coming from their mutual repulsion, whereas the near-degeneracy correlation is taken into account by having more than one determinant. We optimize the Jastrow parameters and determinant coefficients by minimizing the variance of the local energy \cite{Umrigar88PRL}. Finally, we use fixed-node diffusion Monte Carlo (DMC) \cite{Foulkes,Umrigar93JCP} to project the optimized many-body wave function onto a better approximation of the true ground state, an approximation that has the same nodes as $\Psi_T$. 

The fixed-node DMC energy is an upper bound to the true energy and depends
only on the nodes of the trial wave function obtained from VMC. We have
calculated the energy $E(N,L,S)$ of a circular quantum dot for each $N$ with
angular momentum $L$ and spin $S$ -- all good quantum numbers for our model.
($S_z$ is also a good quantum number, and all our calculations are done
for $S_z=S$; however, $E$ is independent of $S_z$) We investigated all possible combinations of $L$ and $S$ for the low lying states, and the combination yielding the lowest DMC energy, $E_{\rm G}$, was taken as the ground state for that $N$. For expectation values of operators that do not commute with the Hamiltonian -- the density or the pair-density, for example -- we use an extrapolated estimator \cite{foot1,Foulkes} (denoted $F_{\rm QMC}$ for an operator $F$) which eliminates the inaccuracy coming from the first-order error in the trial wave function. $F_{\rm QMC}$ is defined as $2 F_{\rm DMC} - F_{\rm VMC}$ when $F_{\rm DMC} \geq F_{\rm VMC}$ and as $F_{\rm DMC}^2/F_{\rm VMC}$ otherwise.

In the multi-determinant expansion of $\Psi_T$, we keep only Slater determinants formed from the lowest energy Kohn-Sham orbitals for all of the results shown here. Our study is currently limited to $r_s \le 18$ for technical reasons. The most serious is the failure of the VMC optimization as many Slater determinants need to be included for stronger interactions.

For two cases, corresponding to one moderate and one large $r_s$, we have done preliminary calculations with higher orbitals by including all determinants involving promotion of 2 electrons across a shell gap (10 configuration state functions for $N=20$). This, then, allows for a change in the nodes of $\Psi_T$. We find that the change in the energy, as well as the change in density and pair-density, is small, though somewhat larger for greater $r_s$. Thus we believe that the fixed-node error in our calculations is under control.

\textit{Acknowledgments ---} This work was supported in part by the NSF
(grants DMR-0506953 and DMR-0205328). AG was supported in part by the funds
from the David Saxon chair at UCLA.





\end{document}